\documentclass[conference]{IEEEtran}
\IEEEoverridecommandlockouts
\usepackage{cite}
\usepackage{amsmath,amssymb,amsfonts}
\usepackage{algorithmic}
\usepackage{graphicx}
\usepackage{textcomp}
\usepackage{xcolor}
\usepackage[acronym]{glossaries}  
\usepackage{url}
\usepackage{breakurl}

\def\BibTeX{{\rm B\kern-.05em{\sc i\kern-.025em b}\kern-.08em
    T\kern-.1667em\lower.7ex\hbox{E}\kern-.125emX}}
\newacronym{ai}{AI}{Artificial Intelligence}
\newacronym{unr155}{UNR155}{United Nations Regulation No. 155}
\newacronym{unr156}{UNR156}{United Nations Regulation No. 156}
\newacronym{csms}{CSMS}{Cyber Security Management System}
\newacronym{sums}{SUMS}{Software Update Management System}
\newacronym{ota}{OTA}{Over-The-Air}
\newacronym{nis}{NIS2}{Network and Information Systems Directive}
\newacronym{nhtsa}{NHTSA}{National Highway Traffic Safety Administration}
\newacronym{cartac}{CARTAC}{China Automotive Technology and Research Center Co., Ltd}
\newacronym{tbt}{TBT}{Technical Barrier to Trade}
\newacronym{ccc}{CCC}{China Compulsory Certification}
\newacronym{ecu}{ECU}{Electronic Control Module}
\newacronym{dssad}{DSSAD}{Data Storage System for Automated Driving}

\newacronym{oem}{OEM}{Original Equipment Manufacturer}
\newacronym{iso21434}{ISO-21434}{Road vehicles — Cybersecurity engineering}
\newacronym{iso24089}{ISO-24089}{Road vehicles — Software update engineering}
\newacronym{cra}{CRA}{Cyber Resilience Act}

\newacronym{tara}{TARA}{Threat Analysis and Risk Assessment}

\begin{document}

\title{Navigating the road to automotive cybersecurity compliance

\thanks{This work was supported by project SERICS (PE00000014) under the MUR National Recovery and Resilience Plan funded by the European Union - NextGenerationEU}
}

\author{\IEEEauthorblockN{Franco Oberti\textsuperscript{1,2}\thanks{Authors contacts: \{franco.oberti, alessandro.savino, Fabrizio Abrate, stefano.dicarlo\}@polito.it and filippo.parisi@dumarey.com}, Fabrizio Abrate\textsuperscript{3}, Alessandro Savino\textsuperscript{1}, Filippo Parisi\textsuperscript{2}, and Stefano Di Carlo\textsuperscript{1}}
\IEEEauthorblockA{\textsuperscript{1}\textit{Control and Computer Eng. Dep., Politecnico di Torino}
Torino, Italy \\
\textsuperscript{2}\textit{Dumarey Softronix}, Torino, Italy}
\textsuperscript{3}\textit{Dumarey Automotive}, Torino, Italy}


\maketitle

 \IEEEpubidadjcol

\begin{abstract}

Modern vehicles are now part of a complex digital ecosystem, leveraging \gls{ai} and cloud computing for enhanced safety, efficiency, and user experience. However, this digital integration has introduced significant cybersecurity challenges, including data protection, unauthorized access prevention, and user privacy.

As vehicles become more vulnerable to cyber-attacks, the industry must implement robust cybersecurity measures. Regulations like the UN's UNR155 and UNR156 establish stringent cybersecurity requirements, demanding comprehensive management systems, regular updates, and continuous testing to counter evolving threats. These regulations underscore the importance of cybersecurity in automotive safety.

Future automotive cybersecurity will depend on developing advanced protections and collaboration among manufacturers, policymakers, and cybersecurity experts to ensure innovation and security in an interconnected digital world.
\end{abstract}


\section{Introduction}

The automotive industry, which began in 1908 with the introduction of the Ford Model T, has undergone profound transformations \cite{alyass2024,battista2024}. Today's vehicles are more than mechanical marvels; they incorporate advanced connectivity features that link them to a broad digital ecosystem. This evolution has significantly improved safety, efficiency, and the overall driving experience \cite{perner2024}. However, integrating these technologies introduces new challenges, particularly in cybersecurity.

Modern vehicles boast advanced connectivity, incorporating \gls{ai} \cite{jamal_impact_ai_2024} and cloud computing technologies \cite{sysgo_edge_cloud_2024}. These features enable vehicles, intelligent infrastructures, and digital systems to communicate, enhancing safety, efficiency, and driving experience.  Conversely, vehicles become more integrated and connected and more susceptible to cyber-attacks \cite{salesforce_security_2023,zheng2024,kong2024,taherdoost2024}. Major cybersecurity challenges include securing data, protecting vehicles, and ensuring privacy \cite{karahasanovic2017,wirelesscar_challenges_2023,costantino2024,moore_auto_osint,burkacky2020,singh2021}.


The fully connected automotive world offers immense benefits but poses significant cybersecurity risks \cite{kim2020}. To ensure these risks do not undermine the benefits, the automotive industry must implement comprehensive and robust cybersecurity strategies \cite{singh2021}. By prioritizing security and fostering collaboration among stakeholders\cite{sommer2019,nilsson2008}, the industry can protect its advancements and continue to innovate securely and effectively. Accordingly, governments and companies have issued numerous regulations and engineering standards focusing on cybersecurity in automotive technology. These standards emphasize continuous innovation and vigilance, stakeholder collaboration, regular updates and testing, and proactive threat mitigation. This paper overviews the regulative effort to address these challenges and the essential role of ongoing compliance to adapt to evolving threats in the automotive sector.

\section{Automotive Cybersecurity Legislation}

In Europe and partner markets such as Korea and Japan, evolving regulations focus on cybersecurity for road vehicles. Two crucial regulations are \gls{unr155} and \gls{unr156}.

\subsection{United Nations Regulation No. 155}
\gls{unr155} \cite{unr155_2021} addresses cybersecurity management systems for vehicles, aiming to mitigate the increasing threats of cyber attacks in the automotive sector. Adopted by the United Nations Economic Commission for Europe (UNECE), it became effective in January 2021. This regulation mandates that all new vehicle types must comply starting from July 2022, and all vehicles produced must comply by July 2024. The Technical Assessment (TA) process under \gls{unr155} involves a thorough evaluation of a manufacturer's \gls{csms} by an independent authority or recognized technical service. The \gls{csms} must demonstrate the manufacturer's capability to identify, manage, and mitigate cyber risks throughout the vehicle's lifecycle, focusing on threat detection, incident response, and security measures implementation during the design, production, and post-production stages. A crucial requirement is that the \gls{csms} certification is valid for three years, after which re-certification is necessary to ensure continuous compliance and adaptation to evolving cyber threats.

\subsection{United Nations Regulation No. 156}\label{subsec:UNR}
\gls{unr156} \cite{unr156_2021} complements \gls{unr155} by focusing on software updates and software integrity within vehicles. It ensures that vehicles remain secure and function correctly throughout their operational life, especially when receiving software updates. Key requirements include the deployment of \gls{sums}, secure \gls{ota} update processes, and continuous monitoring and validation of software integrity. While both \gls{unr155} and \gls{unr156} emphasize security, \gls{unr156} distinctly focuses on the lifecycle management of vehicle software.

\subsection{Network and Information Systems Directive}\label{subsec:NIS}
The \gls{nis}  \cite{nis_2018} expands the scope of cybersecurity beyond just the vehicle to include the infrastructure supporting vehicle manufacturing and services. It aims to enhance the resilience of critical infrastructures within the automotive sector. Key measures include enhanced cybersecurity protocols for industrial control systems, improved incident response capabilities, and mandatory reporting of significant cyber incidents. \gls{nis} affects a broader spectrum of the automotive industry, targeting supply chains, manufacturing plants, and service networks.

The introduction of \gls{unr155}, \gls{unr156}, and \gls{nis} represents a significant advancement in securing the automotive industry against emerging threats. These regional regulations enhance vehicle and infrastructure security and set a precedent for regional regulatory practices. As the industry adapts to these changes, ongoing collaboration and innovation will be crucial to meet the evolving demands of regulatory compliance.

\begin{figure}[htb]
    \centering
    \includegraphics[width=0.49\textwidth]{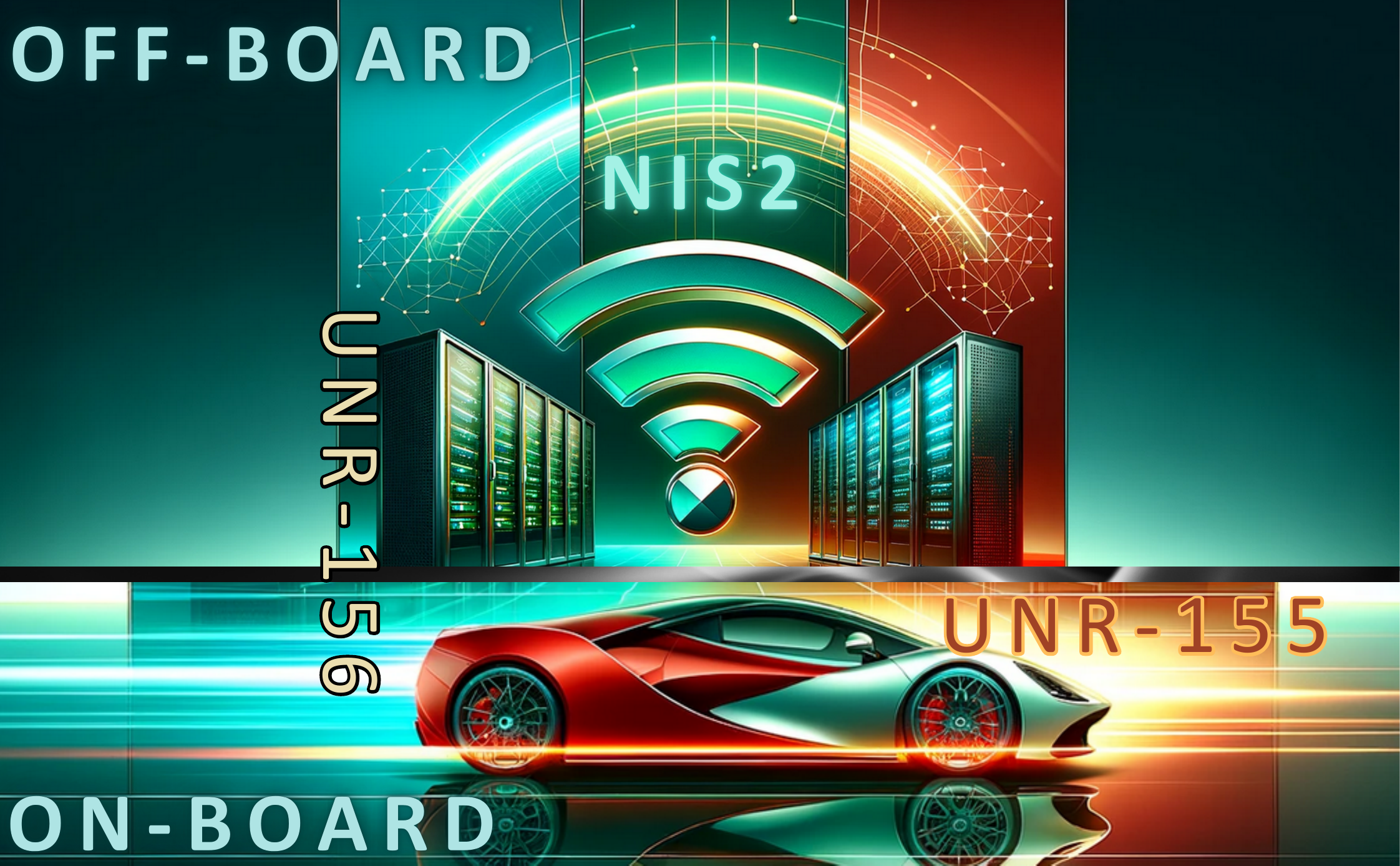}
    \caption{The figure illustrates the cybersecurity focus areas of UNR-155, NIS2, and UNR-156 within the context of vehicle functionality. UNR-155 primarily covers cybersecurity in the onboard domain, directly addressing the product's security. NIS2 operates in the off-board domain, handling cybersecurity related to the IT infrastructure that supports vehicle functionality. UNR-156 acts across both domains, providing comprehensive cybersecurity coverage for both the onboard systems and the supporting off-board IT infrastructure}
    \label{fig:ONFBOARD}
\end{figure}

In the U.S., the \gls{nhtsa} \cite{nhtsa2022} has released an updated version of ``Cybersecurity Best Practices for the Safety of Modern Vehicles,'' improving upon its 2016 edition. This document offers guidance for the automotive industry to enhance vehicle cybersecurity, ensuring safety.

According to the \gls{nhtsa}, as vehicle technology and connectivity continue to advance, maintaining a strong focus on cybersecurity is imperative for automakers, developers, and operators. The \gls{nhtsa} remains committed to ensuring the safety of vehicles on the nation's roads, providing updated best practices to help the industry mitigate cybersecurity risks.

Agency research, industry voluntary standards, and recent insights from motor vehicle cybersecurity studies inform the 2022 Cybersecurity Best Practices. Reflecting public feedback received on the draft published in the Federal Register in 2021, this non-binding document encapsulates critical best practices anticipated to significantly influence the industry. The \gls{nhtsa} routinely assesses cybersecurity risks and best practices, planning to update these guidelines to address the ongoing evolution of motor vehicles and their cybersecurity technologies.

In September 2023, the Technical Committee on Intelligent and Connected Vehicles under \gls{cartac} \cite{cartac2023} completed a technical review of two crucial standards: ``Technical Requirements for Vehicle Cybersecurity'' and ``Intelligent and Connected Vehicle - Data Storage System for Automated Driving.'' These standards are poised to proceed through \gls{tbt} notification stages and are expected to be officially released in the second quarter of 2024. It is anticipated that the implementation of ``Technical Requirements for Vehicle Cybersecurity'' will occur concurrently with the ``General Technical Requirements for Software Update of Vehicles.''

The three mandatory standards discussed—pertaining to automotive software updates, cybersecurity, and data storage systems for automated vehicles—are aligned with UN R155, R156, and the draft \gls{dssad}. As of October 2023, all have successfully passed technical review.

\subsection{Introduction to the Formulation of Chinese Mandatory Standards}

To clarify the process of formulating mandatory standards in China, particularly for ``Technical Requirements for Vehicle Cybersecurity,'' we outline the development timeline following the general procedures in the Management Measures for Mandatory National Standards. Completing the technical review by the committee confirms that the technical specifications of the standard have been finalized, setting the stage for a predicted official release in the second quarter of 2024 based on the timelines of previous standards.

\subsection{Estimated Implementation Date of the Standard}

The standards ``Technical Requirements for Vehicle Cybersecurity'' and ``Intelligent and Connected Vehicle - Data Storage System for Automated Driving'' are expected to be released in May 2024 and to come into force in November 2024. As per the adoption procedure into the \gls{ccc} rules, Technical Committee TC 114 will evaluate these standards and determine their inclusion timeline into the \gls{ccc}. Based on recommendations from the working group, the implementation timeline is expected to be as follows:
\begin{itemize}
\item Release in May 2024
\item Enter into force in November 2024
\item From January 2025, the standards will be integrated into implementation rules for \gls{ccc} and will apply to new vehicle models
\item From January 2026, the standards will apply to all vehicle models
\end{itemize}

\subsection{Coordination with CCC Implementation}

Further regulations are necessary to coordinate the implementation of these standards within the \gls{ccc} framework. Critical considerations include evaluating compliance certificates for \gls{csms} and \gls{sums} with UN R155 and UN R156, as well as the preparations and resources required for validation testing. Ongoing monitoring of the development of supportive rules is essential.

\section{Emerging Challenges under European Regulation}

\gls{unr155} represents a significant advancement in automotive cybersecurity, establishing stringent requirements that manufacturers must adhere to. With a phased implementation schedule and mandatory re-certification every three years \cite{eurlex2021,sandler2022}, this regulation ensures that new vehicles will be better equipped to handle cyber threats, fostering a safer and more secure automotive ecosystem. 

One of the primary challenges for \gls{oem}s under \gls{unr155} is communicating cybersecurity needs and instilling a cybersecurity mindset across the entire supply chain. This complex task demands a comprehensive approach to ensure all parties comprehend and implement cybersecurity measures. In this context, \gls{iso21434} \cite{iso21434} plays a crucial role in supporting \gls{oem}s by promoting cybersecurity awareness and practices. While \gls{iso21434} is an engineering standard and \gls{unr155} is a regulation, designing systems according to the first one provides confidence that they follow the principles that align with the second one's requirements. A similar supportive role is fulfilled by \gls{iso24089} \cite{iso24089}, which also complements the framework established by \gls{unr155}. These standards collectively form a robust structure for managing automotive cybersecurity.

Despite these comprehensive frameworks, significant challenges persist in the vulnerability disclosure process. When vulnerabilities are identified, it is critical to efficiently disseminate accurate information to \gls{oem}s and suppliers to enable them to assess the impact and severity of the threat. However, this process is hindered because the complete vehicle architecture often remains undisclosed, and \gls{tara} details are not uniformly shared across different entities.

Even though cybersecurity regulations in the automotive domain have been issued recently, ongoing discussions and proposals about changes, updates, and new topics continue. Some automotive suppliers are keen on obtaining a certification of organizational processes, responsibilities, and governance to treat risk associated with cyber threats to gain more autonomy and partially relieve \glspl{oem} of their responsibilities through the supply chain. This interest could potentially lead to regulatory updates in the future to accommodate it. Although \gls{unr155} currently applies only to road vehicles (specifically types M and N)  and trailers (specifically type O, in case fitted at least with one ECU), other categories, such as agricultural vehicles, are not covered. Likely, all vehicle categories will eventually be regulated, but it is uncertain whether \gls{unr155} will be the framework used. For instance, the agricultural sector might prefer a different regulatory framework, such as the \gls{cra}, instead of \gls{unr155}. However, vehicle categories M and N, currently governed by \gls{unr155}, are unlikely to be affected by \gls{cra} regulations.

Under \gls{unr155}, \glspl{oem} are solely responsible for the cybersecurity of the products they manufacture. They must also ensure attack resiliency for each subsystem within the product and guarantee that cybersecurity requirements are met throughout the supply chain.

The \gls{cra} is notable because it applies to many products. Under this framework, any manufacturer of systems containing digital elements must perform certification. Therefore, the entity that integrates all subsystems into the final product must ensure that the integration of all certified components remains secure. In a cyber incident, responsibility extends beyond the final manufacturer to include the suppliers.

\begin{figure}[htb]
    \centering
    \includegraphics[width=0.49\textwidth]{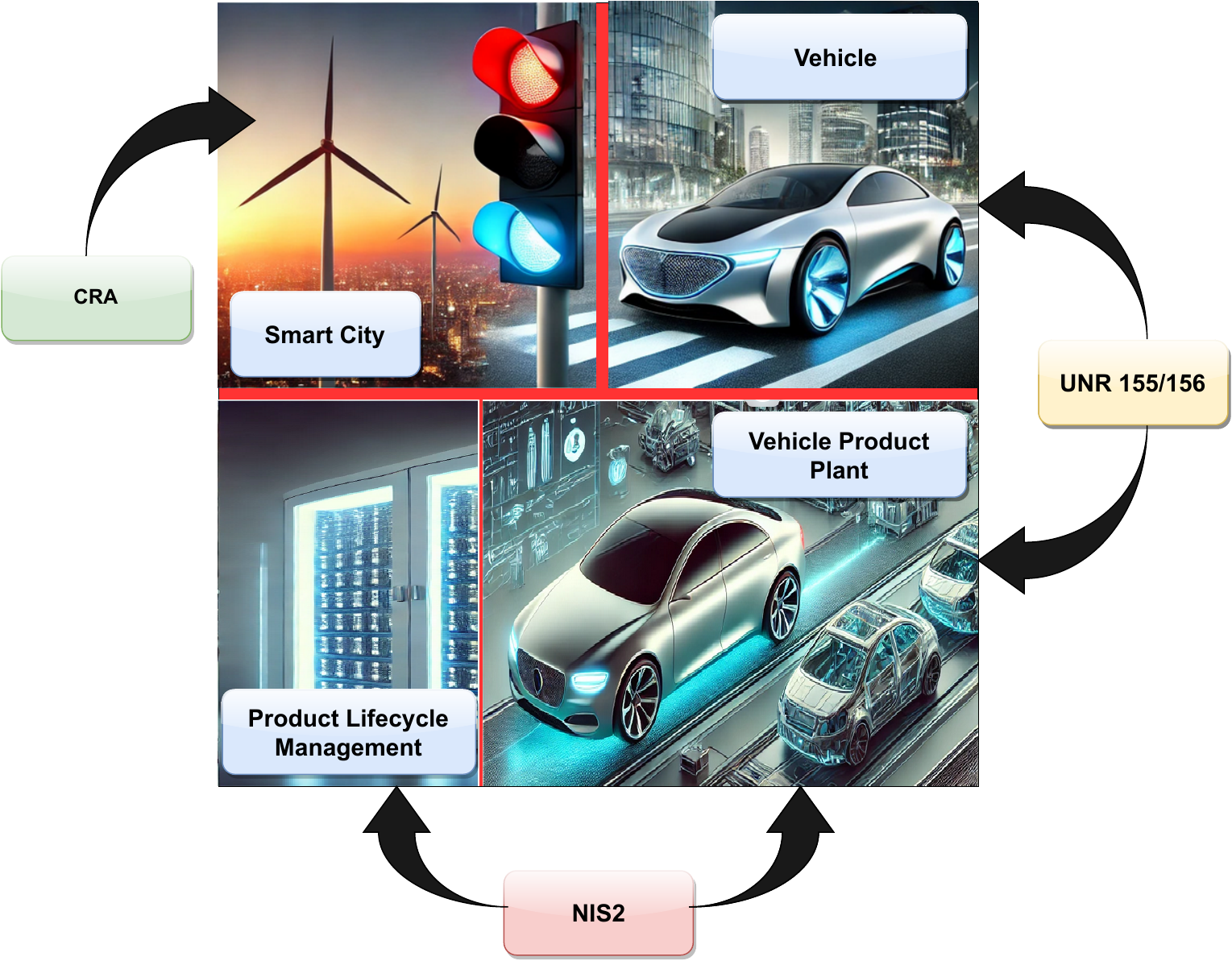}
    \caption{The figure illustrates the impact of cybersecurity regulations (UNR-155, NIS2, UNR-156, and CRA) on the automotive industry and related sectors.}
    \label{fig:cra}
\end{figure}

In situations where a system with digital elements is used in both automotive types M and N, covered under \gls{unr155}, and other domains regulated by the \gls{cra}, certain challenges arise. Since several  \glspl{ecu} are shared between automotive and other domains, these systems may be subject to requirements from both the vehicle manufacturer (under automotive regulations) and the \gls{cra} for other applications (see Figure \ref{fig:cra}). This necessitates managing the same product under two different governance frameworks simultaneously.

\section{Conclusion}

As the automotive industry continues to evolve, cybersecurity is emerging as a fundamental aspect of automotive system value, surpassing even the systems' advanced features. This shift underscores the necessity to integrate cybersecurity from the very inception of the design process, adhering to the 'secure by design' principle. By incorporating security measures at the earliest stages of development, companies can significantly reduce costs and enhance the robustness of their systems.

Moreover, the dynamic nature of cyber threats requires a flexible and adaptive system architecture. This necessitates an agile organizational structure within companies to enable rapid responses to emerging threats. The automotive industry can better protect its assets and sustain consumer trust by fostering an environment that promotes swift adaptability and proactive security measures.

\bibliographystyle{IEEEtran}
\bibliography{biblio}
\vspace{12pt}

\end{document}